\documentclass[aps,prc,twocolumn,preprintnumbers,amsmath,showpacs,floatfix,nofootinbib]{revtex4-1}
\usepackage{graphicx}

\newcommand{\dd}{{\rm d}}

\begin{document}

\preprint{BI-TP 2011/008}

\title{Dynamical evolution of heavy quarkonia in a deconfined medium}

\author{Nicolas Borghini, Cl\'ement Gombeaud}
\affiliation{Fakult\"at f\"ur Physik, Universit\"at Bielefeld, 
  Postfach 100131, D-33501 Bielefeld, Germany}

\date{\today}

\begin{abstract}
  We investigate some consequences of the possibility that heavy quarkonia
  in a quark-gluon plasma possess different (quasi-)bound states, between
  which transitions are possible. 
  In particular, we show that the time-evolution eigenstates in the medium
  are mixtures of the vacuum eigenstates. 
  This leads to abundance ratios of quarkonia that differ from those 
  predicted in statistical models or in the sequential-melting picture.
\end{abstract}


\maketitle

\section{Introduction}
\label{s:Introduction}

Heavy quarkonia provide a convenient testing ground for analytical 
approaches to hadron properties in QCD, both in the vacuum and in the 
presence of a (deconfined) medium~\cite{Brambilla:2010cs}.
In the latter case, the original idea of a direct link between charmonium 
suppression and deconfinement~\cite{Matsui:1986dk} has been refined into 
more involved predictions for the behaviors of the various states with 
rising temperature (for recent reviews see Refs.~\cite{Rapp:2008tf,%
  Kluberg:2009wc,Rapp:2009my}).

Many predictions are formulated in a static picture, most noticeably in 
terms of threshold temperatures, above which a given state is entirely 
``melted'', while it remains intact below.
More dynamical approaches to the dissociation and formation or 
recombination of quarkonia in a medium have been considered in various 
classical kinetic frameworks: 
\`a la Boltzmann~\cite{LevinPlotnik:1995un,Polleri:2003kn,Yan:2006ve}, in
Langevin or Fokker--Planck descriptions~\cite{Patra:2001th,Young:2008he}
or through rate equations~\cite{Grandchamp:2003uw,Zhao:2011cv}.

A common feature of these studies is their focus on charmonia.
This is quite natural, since this corresponds to most of the existing 
experimental results. 
Now, the general expectation is that $c\bar c$ pairs can only exist in a 
quark-gluon plasma (QGP) as either a $J/\psi$ or an unbound system. 
Accordingly, the above mentioned studies consider these two possibilities 
only. 
On the other hand, it is thought that several bottomonia are still bound in
a QGP just above the deconfinement temperature. 
This opens a richer spectrum of possible behaviors for $b\bar b$ pairs, 
especially if transitions can be induced between different bound states.
In this Letter, we wish to take this latter possibility seriously, and 
to investigate some of the consequences of this ansatz, restricting 
ourselves to inner degrees of freedom. 

To this effect, we shall hereafter perform a Gedankenexperiment: 
we put at time $t=0$ a bound $Q\bar Q$ pair in a static, infinitely large 
QGP at temperature $T$.
Then we let the system evolve, and follow the populations of the various 
quarkonium states in time. 

In Section~\ref{s:Model-method} we introduce our model for the heavy 
quarkonia, the QGP and their interaction, as well as for the resulting 
dynamical equations. 
Section~\ref{s:Results} contains our results for the evolving populations
of quarkonium states. 
Further results, in particular involving the external degrees of freedom of
quarkonia, will be presented in a longer, more technical 
publication~\cite{BG_inprep}.
Eventually, we discuss both our model with its underlying assumptions and 
our results in Section~\ref{s:Discussion}.

\section{Model and method}
\label{s:Model-method}

In this Section, we describe our model for the heavy quarkonium states and 
the medium in which they are immersed, as well as for the coupling between 
both.
Then we briefly introduce the equations that govern the dynamics of the 
$Q\bar Q$-state populations. 

\subsection{Quarkonium in a quark-gluon plasma as a \{system + reservoir\}
  dissipative system}
\label{s:Model}

Our purpose in the present study is to investigate the evolution of heavy 
quarkonium states under the influence of the thermal degrees of freedom of 
the plasma, and especially gluons.
Thus, we do not consider (light) quarks, which would interact with the 
heavy quark or antiquark through non-thermal gluons: our medium is a gluon 
plasma. 

We model this plasma as a static bath of harmonic excitations, whose 
frequencies span a continuum $\{\omega_\lambda\}$.
The bath is treated as a thermal reservoir, whose thermodynamic properties
are not affected by the transitions between the various states of the 
embedded quarkonia. 

The modeling of heavy quarkonia is far more delicate than that of the 
plasma. 
In this Letter, we wish to identify generic behaviors that follow from 
allowing quark-antiquark pairs to be in different (quasi-)bound states in a
deconfined medium, between which medium-induced transitions are permitted.
In that view, we deliberately adopt an admittedly simplistic quarkonium 
model, which relies on a minimal number of parameters, deferring more 
realistic modeling to further studies. 
In particular, we make a number of assumptions, which we shall further 
discuss in Section~\ref{s:Discussion}.

Our first model assumption is the existence of a non-relativistic in-medium
quark-antiquark potential~\cite{Digal:2001iu,Wong:2004zr,Arleo:2004ge,%
  Alberico:2006vw,Cabrera:2006wh,Mocsy:2007yj} that admits bound states,
thereby neglecting any imaginary part in the potential~\cite{Laine:2006ns}.
To make the model simple, we consider an attractive Coulomb potential 
$V(r)=-\alpha/r$, with its well-known spectroscopy.\footnote{We 
  take $\alpha=0.4$ for $c\bar c$ pairs and a smaller $\alpha\simeq 0.25$
  for $b\bar b$ pairs, to account for the running of the coupling constant 
  and the smaller size of the ground state.}
As we want to investigate the possible influence of transitions between 
levels, we depart from the exact Coulombic spectroscopy for the states that
are bound in the vacuum: $J/\psi$, $\chi_c$ and $\psi'$ on the one hand, 
$\Upsilon$, $\chi_b$, $\Upsilon'$, $\chi'_b$ and $\Upsilon''$ on the other 
one~\cite{PDG10}.\footnote{For given $S$ and $L$ quantum numbers, we 
  consider for simplicity a single (2$L$+1)-fold degenerated state.}
Those states are modeled with Coulomb wave functions, yet with the same 
energy with respect to the ground 1$S$-state as measured in the vacuum.
This is clearly an approximation, which allows us to lift the degeneracy 
between e.g. 2$S$ and 1$P$ states, and thereby permit single-gluon 
transitions between them. 

Obviously, a single gluon cannot induce a transition between color-neutral 
$Q\bar Q$ states. 
We thus further assume that our spectrum of color-singlet states is 
accompanied by a parallel spectrum of color-octet states, which for the 
sake of simplicity we shall denote similarly to their singlet 
counterparts.

Eventually, we have to model the unbound $Q\bar Q$ pairs, which should 
normally constitute a continuum. 
Consistency with our model for in-medium bound states through a real 
potential implies that the states in that continuum have higher energies 
than the bound states, i.e.\ the latter are {\em always\/} energetically 
favored, which is far from granted.
This is even more a problem with the Coulomb potential, which admits 
arbitrarily large bound states with high principal quantum number --- which
in the static Debye-screening picture would appear as unbound.
To get rid of those states, we fix the {\em dissociation\/} threshold by 
considering bound states of the Coulomb potential as describing unbound 
pairs: 2$S$ and 1$P$ states (resp.\ 3$S$ and 2$P$ states) and the higher 
excited states for $c\bar c$ (resp.\ $b\bar b$) pairs.
A minimal approach to mimic the unbound character of such states consists
in forbidding transitions from them back to the bound ones. 
In our computations, we have fully discarded the scattering solutions of 
the Coulomb potential, and considered two or three levels of 
``dissociated'' states, with at least two states per level, to estimate the
error on our result.  

To leave room for the possible ``recombination'' of heavy quark and 
antiquark into a quarkonium state~\cite{Thews:2000rj}, we also slightly 
modified the model by allowing transitions from the lowest dissociated 
states to bound ones. 
The plots we present in Section~\ref{s:Results} are for results from this 
variant of our model. 
Further plots will be shown elsewhere~\cite{BG_inprep}.

Eventually, we need to specify the coupling between a $Q\bar Q$ pair and 
the plasma. 
We assume {\em dipolar coupling\/}, that is, the gluons only interact 
through their chromoelectric field. 
Incidentally, we need also assume that the Bohr frequencies between 
$Q\bar Q$ states are included in the continuum of bath frequencies, so that
transitions between states can be induced.

\subsection{Evolution of the $Q\bar Q$ states in the plasma}

Now that we have specified the ingredients of our model, we can turn to the
time evolution. 
Further details will be given in a longer publication~\cite{BG_inprep}, 
here we shall merely outline the calculation. 

We expect that the state of the quark-antiquark system at a given time 
should be a statistical superposition of (vacuum) eigenstates. 
Then a natural approach is to use the master-equation formalism~\cite[%
  Chapter 4]{CDG2}. 
Within the decorrelation approximation --- i.e., technically, assuming that
the density matrix of the whole system factorizes into the product of the 
density matrix $\rho^{Q\bar Q}$ of the $Q\bar Q$ pair and that of the 
plasma at every time ---, which amounts to considering an expansion up to 
second order in the coupling potential between the quark-antiquark pair and
the plasma, the {\em populations\/} (diagonal elements) of $\rho^{Q\bar Q}$
are governed by the coupled Einstein equations
\begin{equation}
\label{eq:Einstein}
\frac{\dd\rho^{Q\bar Q}_{ii}}{\dd t}(t) = 
 -\sum_{k\neq i}\Gamma_{i\to k}\rho^{Q\bar Q}_{ii}(t) + 
 \sum_{k\neq i}\Gamma_{k\to i}\rho^{Q\bar Q}_{kk}(t),
\end{equation}
where the transition rates $\Gamma_{i\to k}$ between $Q\bar Q$ levels 
follow from Fermi's golden rule (except for those we set to zero, to mimic
the continuum, as explained above).
These rates involve a sum over the states of the QGP, weighted by their 
respective probabilities. 
This introduces a dependence of all $\Gamma_{i\to k}$ on the plasma 
temperature $T$. 

In the following Section, we present solutions to these evolution equations
for the $c\bar c$ and the richer $b\bar b$ systems.

\section{Results}
\label{s:Results}

The Gedankenexperiment we discussed in Section~\ref{s:Introduction} amounts
to picking out an initial condition at $t=0$ for Eqs.~\eqref{eq:Einstein}, 
for instance $\rho^{Q\bar Q}_{ii}(t\!=\!0)=1$ for the ground 1$S$ state and
$0$ for the excited levels, and to solve the coupled system. 

\begin{figure*}[t!]
  \includegraphics*[width=0.495\linewidth]{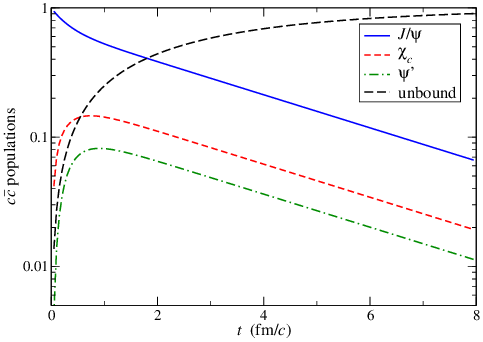}
  \includegraphics*[width=0.495\linewidth]{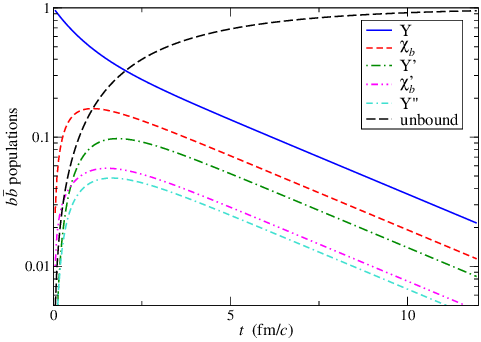}
  \caption{\label{fig:t-dependences}Time dependence of the populations of 
    bound, recombining and unbound states. 
    Left: $c\bar c$ pairs at $T=2T_c$; 
    right: $b\bar b$ pairs at $T=5T_c$.
    Here and in Fig.~\ref{fig:T-dependences}, $T_c=170$~MeV.}
\end{figure*}

A first observation, independent from the initial condition, follows 
directly from the structure of the coupled evolution equations: 
the latter are not diagonal. 
As a consequence, the populations corresponding to the vacuum eigenstates 
of the potential do not constitute an eigenstate of the ``evolution 
operator'' for the vector of populations that can be read off 
Eqs.~\eqref{eq:Einstein}; 
rather, they are linear combinations of the latter. 
That is, the different quarkonium states do not evolve independently from 
each other, but they are coupled together by the medium. 
For instance, even if $\Upsilon'$ is constantly either dissociated or 
decaying into $\chi_b$, at the same time it is recreated, with different 
rates, through the excitation of $\chi_b$ or --- when we allow for that 
possibility --- through recombination of an unbounded $b\bar b$ pair. 
Thus we cannot have the total disappearance of a state in the plasma at 
finite times, as implied by static approaches to quarkonium suppression. 

Let $\Gamma$ denotes the smaller (in absolute value) of the eigenvalues of 
the evolution operator for the populations.  
The medium-induced mixing of states is such that after some transient 
behavior, which depends on the specific choice of initial condition, the
populations $\rho^{Q\bar Q}_{ii}$ of all bound states (and of the unbound 
states that are allowed to recombine) evolve with the same characteristic 
time scale $\Gamma^{-1}$. 
We illustrate this in Fig.~\ref{fig:t-dependences}, which shows the time 
evolution of the populations of $c\bar c$ (left) and $b\bar b$ (right) 
states at respectively $2T_c$ and $5T_c$. 
The characteristic evolution times for these systems at the chosen 
temperatures within our model are respectively 3.4 and 3.8~fm/$c$. 
Quite obviously, for a given system $\Gamma^{-1}$ decreases with rising 
temperature. 

\begin{figure}[b]
  \includegraphics*[width=\linewidth]{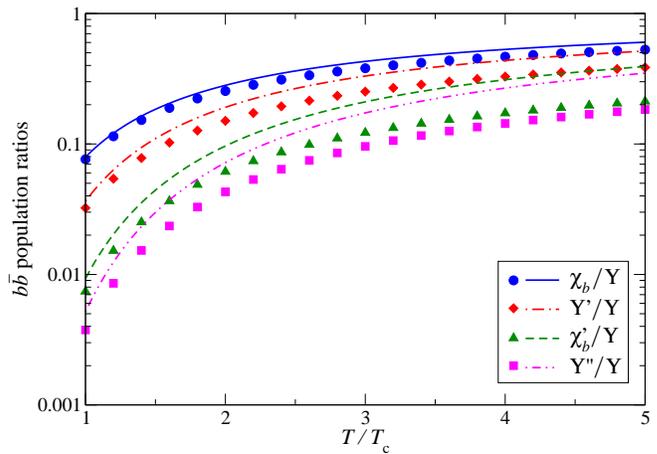}
  \caption{\label{fig:T-dependences}Temperature dependence of the ratios 
    of the populations of excited $b\bar b$ states to that of the ground 
    state ($\Upsilon$) in the stationary regime (see text).
    Curves: corresponding ratios in a thermally equilibrated system.}
\end{figure}
Past the transient regime, the populations of the various bound and 
recombining states are ``equilibrated'' with each other, in the sense that 
the population ratios remain stable. 
(There is however no strict equilibrium, since $Q\bar Q$ pairs are 
consistently lost to non-recombining states). 
These ``stationary'' population ratios depend on the plasma temperature, as
shown for $b\bar b$ pairs in Fig.~\ref{fig:T-dependences}. 
The stationary ratios differ significantly from those for thermally 
equilibrated levels, which is easily understandable. 
The thermal ratios are those which make Eqs.~\eqref{eq:Einstein} stationary
with {\em all\/} transition rates fulfilling the detailed balance condition
\begin{equation}
\label{eq:detailed-balance}
\Gamma_{i\to k}\,{\rm e}^{-E_i/T} = \Gamma_{k\to i}\,{\rm e}^{-E_k/T}\quad
\forall\, i,k.
\end{equation}
Here, the condition is obeyed only by bound ($\Upsilon$, $\chi_b$, 
$\Upsilon'$) and recombining ($\chi'_b$, $\Upsilon''$) states, but 
not by the unbound ones, from which there are no transition. 
It is thus normal that the resulting stationary ratios diverge from the 
thermal ratios. 

We have checked that when we do not suppress back transitions from the 
unbound states, but set them according to condition~\eqref{%
  eq:detailed-balance}, then the stationary abundance ratios are the same 
as in thermal equilibrium.

Given the crudeness of our model, which we wish to further discuss in the
next Section, we have not attempted at this stage to make predictions for 
observables in real nucleus-nucleus collisions, which would anyway 
necessitate some extra modeling of the plasma kinetics as well as 
accounting for the hadronic phase.

\section{Summary and discussion}
\label{s:Discussion}

We have examined the dynamics of the populations of heavy quarkonium states
in a static QGP within the quantum-mechanical master-equation formalism, 
assuming that some elements of the quarkonium spectroscopy survive in the 
plasma, in particular that different bound states exist, between which 
transitions can be induced by the medium. 
Under this assumption, we find that the bound states are mixed together by 
the medium, so that they all evolve with the same characteristic time 
scale. 
This differs from the usual picture of sequential melting, inasmuch as no 
bound state can totally disappear while others would survive. 

In addition, we find that after some transient regime, the population 
ratios remain stationary, at values determined by the plasma temperature, 
yet different from the ratios for quarkonium states in thermal equilibrium.
While the fact that vacuum eigenstates are no longer eigenstates in the 
QGP is model-independent, the actual predictions for the abundance ratios 
depend on the model parameters, and could be used to constrain the latter 
from experimental results. 

Let us now discuss our model.
Its key ingredient in our eyes is the ansatz of medium-induced transitions
--- with a large enough rate compared to the inverse of the plasma lifetime
--- between bound quarkonium states. 
This is the element which couples the evolutions of the various bound 
states together, irrespective of the details of their modeling or of the 
mechanism responsible for the transitions.

The latter is important inasmuch as it selects which states are coupled 
together. 
Here we wanted to consider, in analogy to quantum optics, transitions 
induced by the gauge bosons. 
This choice forced us to invoke some hazy ``spectrum of color-octet 
states'', paralleling that of color-singlets --- which is far from being 
granted, given the repulsive nature of the color-octet channel. 
Instead, we could have conjured non-perturbative effects, like some kind of
soft color interactions~\cite{Edin:1995gi}, to instantly turn color octets
into singlets in the medium. 
Such explanations seem to us to be as disputable as our choice in the 
present context. 
Alternatively, one could think of non-color-exchanging processes, as 
quasi-elastic collisions with off-shell quarks or gluons or scattering of 
photons, provided the latter happen with a sufficient rate.\footnote{See 
  Ref.~\cite{Marasinghe:2011bt}, which appeared while this Letter was being
  finalized, for a study of the photoionization of the $J/\psi$ by large 
  magnetic fields in a QGP.}

If one accepts the possibility of medium-induced transitions between bound 
states, then the models for medium, quarkonia and their interaction that we
have used are purposely the simplest ones one can think of, yet still 
realistic enough to illustrate some plausible phenomena. 
The orders of magnitude we obtain for the typical time scales for the 
evolution of quarkonium populations are reasonable, which justifies our 
choice {\em a posteriori\/}. 
The model allows us to compute transition rates between bound states (which
is the reason why we have kept the Coulomb wave functions although with
``wrong'' energies), while these are not known in more realistic models, 
since such transitions have not been investigated before.
For the dissociation widths of the $1S$ states, one could use the known 
results at leading~\cite{BhanotPeskin} or next-to-leading order~\cite{%
  Park:2007zza}: this would represent an improvement, yet only a partial 
one. 

Note that the dissociation process that we consider, as everyone does, is a
{\em classical\/} process, in which energy is transferred to the $Q\bar Q$ 
pair. 
This is inherent to the description by a real potential similar to the 
vacuum one. 
It might turn out that a better description of the transition from bound 
$Q\bar Q$ state to unbound quark and antiquark in a QGP should 
involve some tunneling through a barrier, as studied in hadronic matter in 
Ref.~\cite{Kharzeev:1995ju}.

The master-equation approach we have used relies on a few assumptions, 
which we shall detail elsewhere~\cite{BG_inprep}.
In short, these amount to assuming --- as is also done in Boltzmann, 
Langevin or Fokker--Planck formalisms --- that the typical time 
scale of 
plasma correlations is small against the characteristic time scale of the 
$Q\bar Q$-plasma interaction, i.e.\ the formalism implicitly rests on a 
``weak-coupling'' assumption. 
We are investigating alternate approaches that do not make use of this 
hypothesis~\cite{DBG_inprep}, yet this seems only feasible at the cost of 
some alternative approximations.

Eventually, we have assumed dipolar coupling between a $Q\bar Q$ pair and 
the plasma, discarding chromomagnetic or quadrupolar and higher order 
chromoelectric couplings. 
This is a large-wavelength approximation --- which might be disputable for 
gluons that should resolve the structure of the bound states --- that can 
be released when using a more realistic model of the quarkonia in the 
plasma.

\section{Acknowledgments}

We gratefully thank J.-P.\ Blaizot, J.-Y.\ Ollitrault and H.\ Satz for 
enlightening discussions. 
C.\ G. acknowledges support from the Deutsche Forschungsgemeinschaft under 
grant GRK 881.

\end{document}